\documentclass[twocolumn,showpacs,preprintnumbers,amsmath,amssymb]{revtex4}
\usepackage{graphicx}
\usepackage{dcolumn}
\usepackage{bm}

\newcommand{\rmi}{{\rm i}}

\begin{document}

\title{Compression of an intensive light pulse in a dense resonant medium \\ and photonic-band-gap structures containing it}

\author{Denis V. Novitsky}
 \email{dvnovitsky@tut.by}
\affiliation{%
B.I. Stepanov Institute of Physics, National Academy of Sciences of
Belarus, \\ Nezavisimosti~Avenue~68, 220072 Minsk, Belarus.
}%

\begin{abstract}
Intensive light pulse interaction with a dense resonant medium is
considered. The possibilities of optical switching and pulse
compression at realistic parameters of the medium are analyzed.
Pulse shape transformation in different photonic band gap structures
containing a dense resonant medium is studied. In particular the
effect of dispersion compensation due to nonlinear interaction with
a medium is reported. The possibility of pulse control with another
pulse is considered in the schemes of co- and counter-propagating
pulses. It is shown that a photonic crystal makes controlling more
effective, at least in the case of co-propagating pulses.
\end{abstract}

\pacs{42.65.Re, 42.65.Pc, 42.70.Qs}

\maketitle

\section{\label{intro}Introduction}

The model of resonant two-level medium is studied actively for many
decades (see, for example, \cite{Krukov, Poluektov}). This interest
is caused, in many respects, by the possibility of ultrashort pulses
generation which can be used in different fundamental and practical
applications. The concept of the so-called dense resonant media
implies the necessity of taking into account the interaction of an
atom with local field produced by all other particles of the system.
In other words, one has to consider near dipole-dipole (NDD)
interactions which are characterized by the quantity
\begin{eqnarray}
b=\frac{4 \pi \mu^2 C}{3 \hbar \gamma_2}, \label{b}
\end{eqnarray}
where $\mu$ is the transition dipole moment, $C$ is the volume
density of two-level atoms, $\gamma_2=1/T_2$ is the rate of
transverse relaxation, $\hbar$ is the Planck constant. Nonlinear
effects connected with NDD interactions become noticeable when the
medium is dense enough. One of them, the effect of intrinsic optical
bistability (IOB), was studied in details \cite{Hopf, Malyshev,
Afan98, Brunel}. It was shown that IOB occurs when $b$ is greater
than 4 (in the case of thin films \cite{Friedberg}) or even less (if
we consider propagation effects in extended medium \cite{Novitsky}).
This phenomenon takes place in the steady state, while in pulse
regime one can observe optical switching \cite{Cren92, Scalora} and
soliton formation \cite{Bowd91, Afan02}. In the present paper we
also consider intensive light pulses, so that change of the
population difference (or inversion) cannot be neglected. To study
pulse propagation, the numerical solution of the Maxwell wave
equation was implemented. This allows to automatically take into
consideration the processes of dispersion and diffraction which are
usually ignored in the analytical approaches.

The main attention in this research is devoted to discussion of the
properties of a combination of a one-dimensional photonic crystal
and a dense resonant medium under pulse operation. The stationary
bistable response of such system was analyzed previously
\cite{Novitsky}. It turned out that one can change and control such
characteristics of bistability as hysteresis loop width and
switching intensity by using photonic crystal. In the present paper
pulse form transformation in such nonlinear photonic band gap
structures is studied in comparison with the behavior in linear
case. In particular, possibilities to compensate dispersion
spreading in such system and control pulse properties by using
another pulse are expected to be found.

The paper is divided in several sections. In Section \ref{basic} the
main expressions for description of the system considered are given,
just as brief characterization of the numerical approach used. This
methodology is then applied to obtain the results of the other
sections. Section \ref{coher} is devoted to optical switching and
pulse shape change (compression and splitting) under propagation in
a finite layer of a dense resonant medium. In particular, the role
of near dipole-dipole interactions is estimated. Section \ref{pbg}
contains the calculation results for a single pulse in a nonlinear
photonic crystal, such as possibility of dispersive spreading
compensation. Finally, in Section \ref{2pulse} two schemes of
controlling pulse intensity with another pulse are considered.
Photonic crystal is studied as an element which makes this process
more effective, at least in the scheme of co-propagating pulses.

\section{\label{basic}Basic equations and numerical approach}

Let us consider radiation propagation in a dense resonant medium in
$z$-direction. Light interaction with resonant medium with taking
into account nonlinear contribution due to near dipole-dipole
interactions is described by the modified Maxwell-Bloch system as
follows \cite{Bowd93, Cren96}:
\begin{eqnarray}
\frac{dP}{dt}&=&\frac{\rmi \mu}{\hbar} E N + \rmi P \left(\Delta
\omega + \frac{4 \pi \mu^2 C}{3 \hbar} N \right) - \gamma_2 P,
\label{dPdt} \\
\frac{dN}{dt}&=&2 \frac{\rmi \mu}{\hbar} \left(E^* P - P^* E \right)
-\gamma_1 (N - 1), \label{dNdt} \\
\frac{\partial^2 \Sigma}{\partial z^2}&-&\frac{1}{c^2}
\frac{\partial^2 \varepsilon_{bg} \Sigma}{\partial t^2} = \frac{4
\pi}{c^2} \frac{\partial^2 P_{nl}}{\partial t^2}, \label{Max}
\end{eqnarray}
where $N$ is the population difference, $P$ is the microscopic
(atomic) polarization; $\Delta \omega$ is the detuning of the field
frequency $\omega$ from atomic resonance; $\gamma_1=1/T_1$ is the
rate of longitudinal relaxation; $c$ is the light speed in vacuum.
Macroscopic electric field $\Sigma$ is expressed via its amplitude
$E$ as $\Sigma=E \exp[-\rmi(\omega t-kz)]$; similarly for
macroscopic nonlinear polarization we have $P_{nl}=\mu CP
\exp[-\rmi(\omega t-kz)]$. Here $k=\omega/c$ is the wavenumber, and
$\varepsilon_{bg}$ is the background dielectric permittivity,
assumed to be linear and dispersionless.

The system (\ref{dPdt}-\ref{Max}) can be represented in the
dimensionless form by introducing new arguments $\tau=\omega t$ and
$\xi=kz$:
\begin{eqnarray}
\frac{dP}{d\tau}&=&\rmi \Omega N + \rmi P (\delta+\epsilon N) - \tilde\gamma_2 P, \label{dPdtau} \\
\frac{dN}{d\tau}&=&2 \rmi (\Omega^* P - P^* \Omega) -
\tilde\gamma_1 (N-1), \label{dNdtau} \\
\frac{\partial^2 \Omega}{\partial \xi^2}&-&\varepsilon_{bg}
\frac{\partial^2 \Omega}{\partial \tau^2}+2 \rmi \frac{\partial
\Omega}{\partial \xi}+2 \rmi \varepsilon_{bg} \frac{\partial
\Omega}{\partial \tau}+(\varepsilon_{bg}-1)
\Omega \nonumber \\
&&=3 \epsilon \left(\frac{\partial^2 P}{\partial \tau^2}-2 \rmi
\frac{\partial P}{\partial \tau}-P\right), \label{Maxdl}
\end{eqnarray}
where $\Omega=(\mu/\hbar\omega)E$ is the dimensionless amplitude of
electric field (normalized Rabi frequency);
$\delta=\Delta\omega/\omega$ is the normalized frequency detuning;
$\epsilon=4 \pi \mu^2 C/3 \hbar \omega=\tilde\gamma_2 b$ is the NDD
interaction constant which provides extra nonlinearity in Eq.
(\ref{dPdtau}); $\tilde\gamma_j=\gamma_j/\omega$, $j=1, 2$.

Numerical solving of the system (\ref{dPdtau}-\ref{Maxdl}) is
performed using the finite-difference time-domain (FDTD) method. The
explicit scheme to solve Eq. (\ref{Maxdl}) on the computational mesh
$(l\Delta\tau, j\Delta\xi)$ is given by
\begin{eqnarray}
\Omega_j^{l+1}=[-a_1 \Omega_j^{l-1}+b_1 \Omega_{j+1}^l+b_2
\Omega_{j-1}^l+f \Omega_j^l-R]/a_2, \label{explOm}
\end{eqnarray}
where
\begin{eqnarray}
a_1&=&\varepsilon_j(1+\rmi\Delta\tau), \qquad
a_2=\varepsilon_j(1-\rmi\Delta\tau), \nonumber \\
b_1&=&\left(\frac{\Delta\tau}{\Delta\xi}\right)^2 (1+\rmi\Delta\xi),
\qquad
b_2=\left(\frac{\Delta\tau}{\Delta\xi}\right)^2 (1-\rmi\Delta\xi), \nonumber \\
f&=&2\varepsilon_j-2\left(\frac{\Delta\tau}{\Delta\xi}\right)^2+\Delta\tau^2(\varepsilon_j-1);
\nonumber
\end{eqnarray}
$\varepsilon_j$ is the value of dielectric permittivity
$\varepsilon_{bg}$ at $\xi_j=j\Delta\xi$. In Eq. (\ref{explOm}) $R$
stands for the finite-difference representation of the right-hand
member of Eq. (\ref{Maxdl}) which describes nonlinear properties of
the medium. For the case of the dense resonant medium it is
\begin{eqnarray}
R=3\epsilon
\left(P_j^{l+1}(1-\rmi\Delta\tau)+P_j^{l-1}(1+\rmi\Delta\tau)-P_j^l(2+\Delta\tau^2)
\right). \nonumber
\end{eqnarray}

Polarization in the mesh points is obtained from Eqs. (\ref{dPdtau})
and (\ref{dNdtau}). To solve them the well-known midpoint
trapezoidal method is used. To set the boundary conditions we apply
the TF/SF approach, when the full calculation region is divided into
two subregions containing total and scattered fields respectively,
and the so-called absorbing boundary conditions according to the
perfectly matched layer (PML) method. This approach helps to avoid
nonphysical reflections of the scattered field back into the
calculation region \cite{Anantha}.

The calculation scheme considered is similar to that of Ref.
\cite{Cren96} where it was used to analyze validity of the
quasiadiabatic approximation for consideration of ultrashort light
pulses propagation in a dense resonant medium.

\section{\label{coher}Coherent pulse interaction with a dense resonant medium}

\begin{figure}
\includegraphics[scale=0.75, clip=]{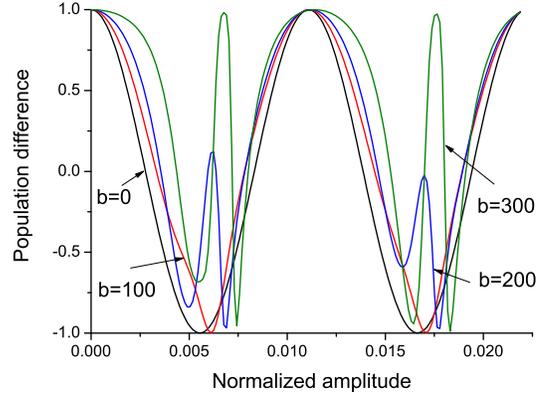}
\caption{\label{fig1} (Color online) Population difference on the
entrance of the layer of the dense resonant medium after pulse
passage versus its amplitude $\Omega_0$ at different values of NDD
interaction constant. Layer thickness $L=\lambda$.}
\end{figure}

As it is known, the cases of coherent and incoherent interactions
are distinguished by the comparison of the pulse duration $t_p$ with
the characteristic relaxation times of the medium $T_1$ and $T_2$,
moreover, as a rule, $T_1>T_2$. Here we consider the coherent case
when $t_p<<T_2<T_1$. In this section the values of parameters are
assumed to be as follows: light wavelength $\lambda=0.5$ $\mu$m,
$t_p=30$ fs, $T_1=1000$ ps, $T_2=100$ ps, $\delta=0$, and, if not
stated another, $\varepsilon_{bg}=1$. The pulse on the entrance of
the medium has Gaussian shape $E=E_0\exp(-t^2/2t_p^2)$. Obviously,
in time intervals comparable with the pulse duration (of the order
of 0.1 ps) the processes of incoherent relaxation can be neglected.
Under these conditions the system consisting of two-level atoms can
demonstrate optical switching from the ground state to the excited
one for the time of the order of the pulse duration \cite{Cren92}.
However, as the authors of Ref. \cite{Scalora} note, taking into
account for the propagation effects (hence, self-phase modulation as
well) not only changes the characteristics of switching, but also
makes it difficult to qualitatively predict them without execution
of rigorous numerical simulations in every particular case.

The calculation results in Fig. \ref{fig1} show that, at small
values of the NDD interaction constant $b$, the dependance of
population difference on the pulse amplitude has periodic form. When
the parameter of extra nonlinearity $b$ is increased, this strictly
periodic situation is disturbed, inversion maximum being shifted
towards greater amplitudes and even being reduced, requiring more
accurate adjustment of pulse intensity. In addition, there is
similar dependance on the layer thickness: strictly periodic
behavior of population difference occurs only at small (as compared
with wavelength) medium thickness.

The difficulties of optical switching, noted in Ref. \cite{Scalora},
are connected with the features considered above, as they have
analyzed the case of large values of NDD interaction constant. In
our designation it corresponds to the values of $b$ of the order of
hundreds and thousands. It seems not to be realistic: usually $b$
does not exceed several units. For example, for gaseous media with
typical parameters $\mu^2=10^{-38}$ erg cm$^3$, $\gamma_2=10^9$
s$^{-1}$, $C=10^{20}$ cm$^{-3}$ \cite{Afan98, Afan99} one obtains
$b\approx4$. In the case of excitonic  media possessing
substantially greater dipole moment ($\mu^2=10^{-36}$ erg cm$^3$,
$\gamma_2=10^{10}$ s$^{-1}$, $C=10^{18}$ cm$^{-3}$ \cite{Afan02}) we
have only $b\approx0.4$. The dramatic increase of the parameter of
NDD interaction is unlikely to expect. In the present paper we keep
to this restriction, so the periodic behavior corresponding to small
values of $b$ will be always valid.

\begin{figure}
\includegraphics[scale=0.75, clip=]{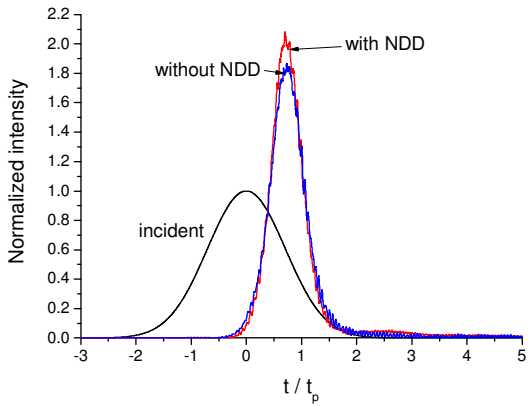}
\caption{\label{fig2} (Color online) Calculation results for pulse
propagation through the layer of the dense resonant medium. The
amplitude of pulse is $\Omega_0=1.5\Delta\Omega_T$. The thickness of
the layer $L=\lambda$; NDD interaction parameter $b=2000$.}
\end{figure}

\begin{figure}
\includegraphics[scale=0.75, clip=]{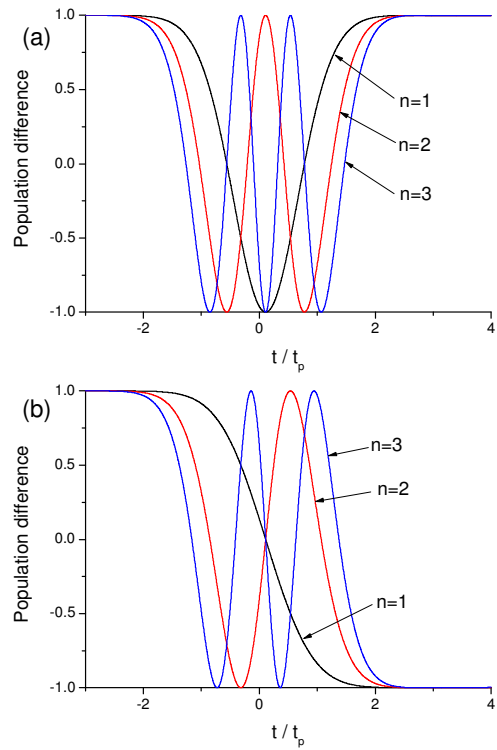}
\caption{\label{fig3} (Color online) Temporal behavior of population
difference on the entrance of the layer of the dense resonant medium
at the amplitude of the pulse (a) $\Omega_0=n\Delta\Omega_T$, (b)
$\Omega_0=(n-1/2)\Delta\Omega_T$. Parameter $b=1$.}
\end{figure}

\begin{figure}
\includegraphics[scale=0.75, clip=]{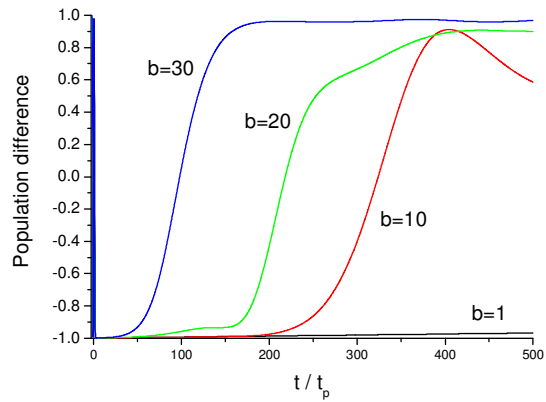}
\caption{\label{fig4} (Color online) Temporal behavior of population
difference on the entrance of the layer of the dense resonant medium
at different values of the parameter $b$. The amplitude of the pulse
is $\Omega_0=1.5\Delta\Omega_T$.}
\end{figure}

\begin{figure}
\includegraphics[scale=0.75, clip=]{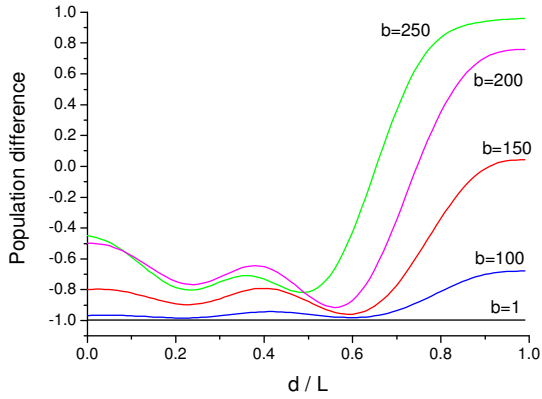}
\caption{\label{fig5} (Color online) Distribution of population
difference along the thickness $L=\lambda$ of the layer of the dense
resonant medium at different values of the NDD interaction
parameter. The amplitude of the pulse is
$\Omega_0=1.5\Delta\Omega_T$.}
\end{figure}

\begin{figure}
\includegraphics[scale=0.75, clip=]{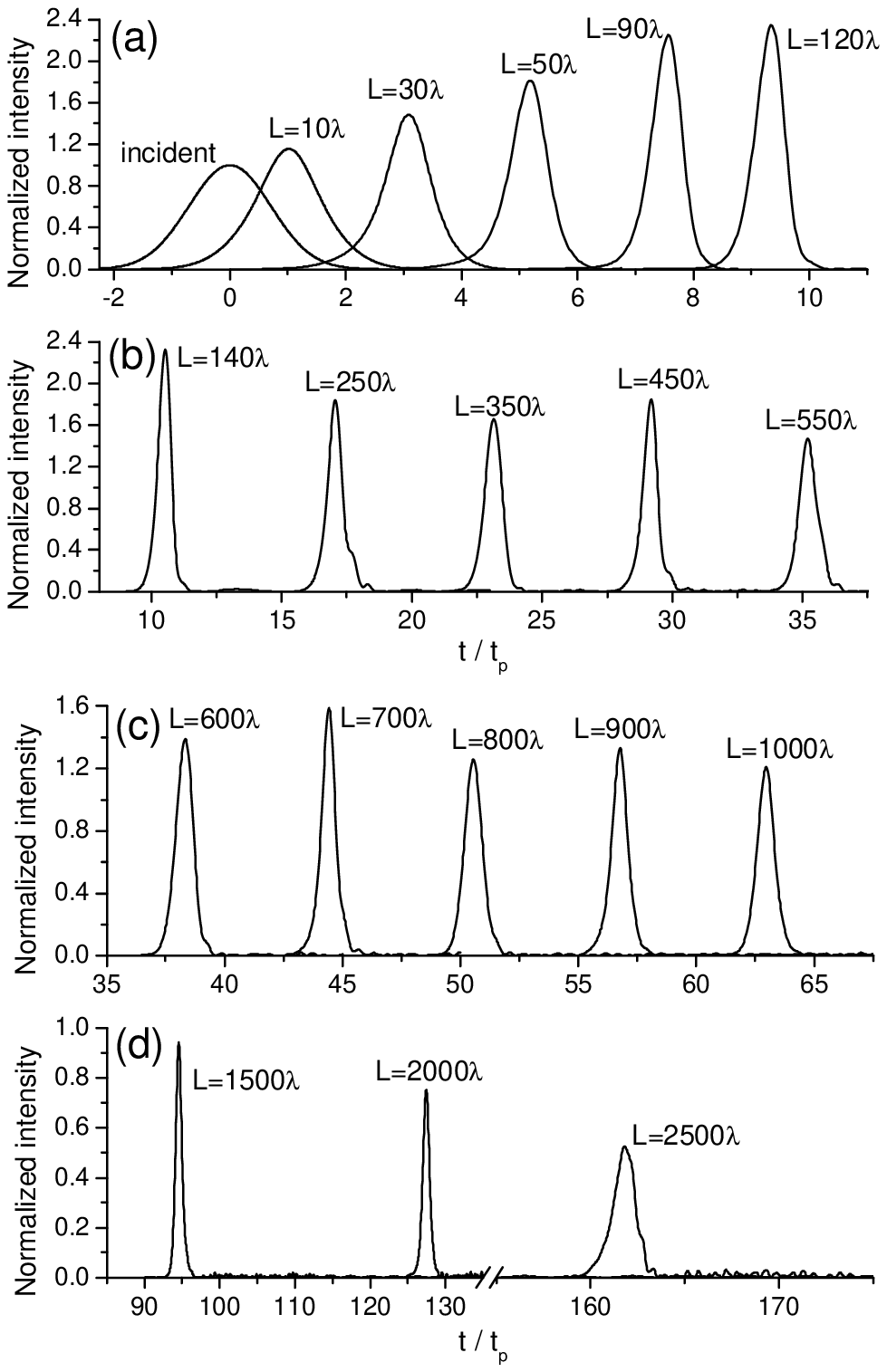}
\caption{\label{fig6} Pulse form transformation after propagation
through the layer of the dense resonant medium of different
thicknesses. The amplitude of pulse is $\Omega_0=1.5\Delta\Omega_T$.
NDD interaction parameter $b=10$.}
\end{figure}

\begin{figure}
\includegraphics[scale=0.75, clip=]{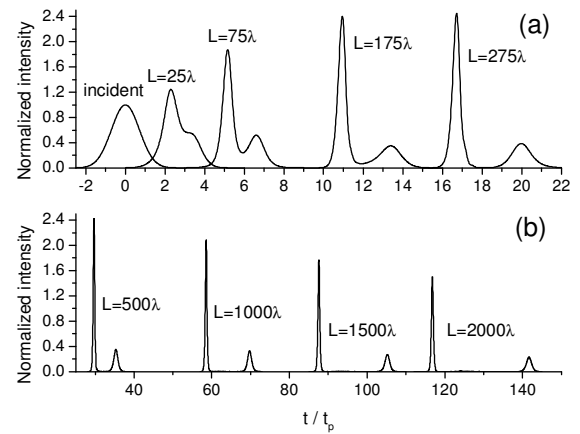}
\caption{\label{fig7} Pulse form transformation after propagation
through the layer of the dense resonant medium of different
thicknesses. The amplitude of pulse is $\Omega_0=2\Delta\Omega_T$.
NDD interaction parameter $b=10$.}
\end{figure}

This implies that the effect of NDD interaction on pulse propagation
in a dense resonant medium is not significant at realistic
parameters. To directly ascertain this conclusion, one should
compare the results of the calculations, when the term including
parameter $b$ is present or absent in Eq. (\ref{dPdt}) or
(\ref{dPdtau}). It turned out that the results remain the same in
both cases. It seems to be obvious, if we recall that the parameter
$b$ amounts to several units (at best, $b=10$). This corresponds to
the utterly small value of $\epsilon\approx10^{-4}-10^{-5}$ in
normalized equation (\ref{dPdtau}). Only when the NDD interaction
constant takes on sufficiently large values, the contribution of
this term becomes essential (Fig. \ref{fig2}). Thus, while in
stationary regime NDD interaction plays vital role in such nonlinear
phenomena as intrinsic optical bistability \cite{Novitsky}, for
coherent pulse propagation this influence seems to be negligible, at
least for realistic values of parameters.

Since the influence of NDD interactions is negligible, one can
estimate the period $\Delta\Omega_T$ of the dependance shown in Fig.
\ref{fig1} by using the commonly known concept of pulse area
\cite{Poluektov}:
\begin{eqnarray}
\vartheta=2\frac{\mu}{\hbar} \int^\infty_{-\infty} Edt. \label{area}
\end{eqnarray}
The period $\Delta\Omega_T$ corresponds to the pulse which returns
atoms after excitation exactly into the ground state. The area of
such pulse should be equal to $2\pi$. Thus, for Gaussian pulse we
obtain
\begin{eqnarray}
\Delta\Omega_T=\frac{\lambda}{2 \sqrt{2\pi}c t_p}. \label{period}
\end{eqnarray}
For the case considered (Fig. \ref{fig1}) it is approximately
$\Delta\Omega_T\approx0.011$. If amplitude of incident pulse amounts
to an integer number of these periods, i.e.
$\Omega_0=n\Delta\Omega_T$, $n=1,2,3,...$, then in temporal behavior
of population difference (Fig. \ref{fig3}a) one can observe $n$
minima and $n-1$ maxima before it achieves a stationary level, in
this instance corresponding to the ground state of the system
($N=1$). In turn, at $\Omega_0=(n-1/2)\Delta\Omega_T$ $n-1$ maxima
and minima occur (Fig. \ref{fig3}b) before the medium becomes
inverted ($N=-1$). However, this final value of population
difference after pulse passage is not stable, as far as relaxation
to the ground state becomes apparent at longer time intervals. It
proceeds the faster, the greater constant $b$ (Fig. \ref{fig4}).
Obviously, incoherent relaxation does not have enough time to
appear. Therefore one can attribute this effect to nonlinear
interaction of radiation with the resonant medium. Moreover, this
relaxation process should be apparent in spatial scale giving rise
to nonuniform distribution of population difference along the layer
thickness (Fig. \ref{fig5}). The similar behavior also takes place
at smaller values of $b$, but significantly larger thicknesses of
the layer.

Since the influence of NDD interactions is negligible, the formation
of stationary pulses (solitons) of sech-shape can be expected. At
the same time energy conservation, area change and shape
transformation (from Gaussian to hyperbolic secant) may cause the
pulse to become shorter and more intensive. However, this process is
limited by diffraction and dispersion of light: pulse begins to
spread as it propagates in the medium. This fact is the reason of
occurrence of the length of optimal pulse compression. In Fig.
\ref{fig6} it amounts about $120\lambda$. Note that, after reaching
this first maximum, several other are observed: at
$L\approx450\lambda$ (Fig. \ref{fig6}b), $L\approx700\lambda$ and
$L\approx900\lambda$ (Fig. \ref{fig6}c). Finally, at certain
(sufficiently large) distance the pulse exhibits ultimate
attenuation and decay (Fig. \ref{fig6}d).

Another important effect that can be observed in this system is
pulse splitting into several components. The number of components
depends on the area of initial pulse. For example, if the amplitude
of the pulse is $\Omega_0=2\Delta\Omega_T$ (area equals $4\pi$), it
undergoes splitting into two ones (Fig. \ref{fig7}) corresponding to
two $2\pi$-pulses. The second (low-intensive) pulse retards more and
more from the first one as they propagate in the medium.
Simultaneously both these pulses lose energy with distance due to
diffraction and dispersion.

\section{\label{pbg}Single pulse in photonic crystal with a dense resonant medium}

\begin{figure}
\includegraphics[scale=0.75, clip=]{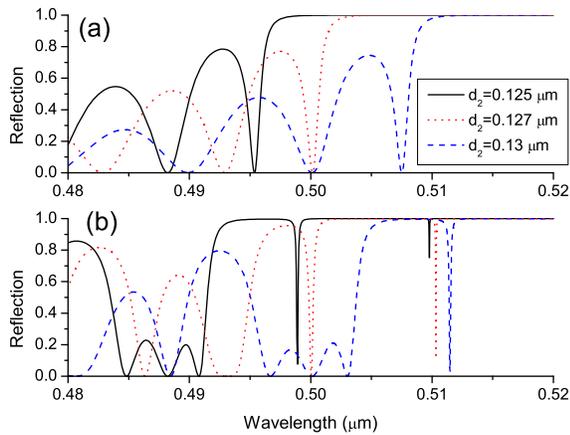}
\caption{\label{fig8} (Color online) Reflection spectra of photonic
crystal (number of periods 8) for different values of the thickness
$d_2$: (a) without a defect, (b) with a defect of $L=20\lambda$ and
$n_3=1$.}
\end{figure}

\begin{figure}
\includegraphics[scale=0.75, clip=]{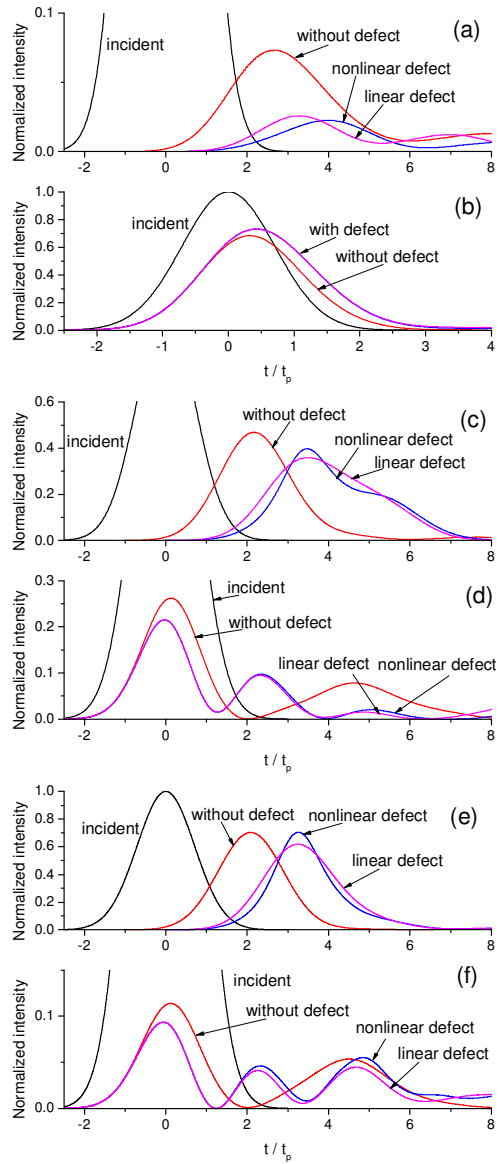}
\caption{\label{fig9} (Color online) The forms of transmitted (a, c,
e) and reflected (b, d, f) pulses after interaction with the
photonic crystal (number of periods 8) with a defect layer of
thickness $L=20\lambda$, $n_3=1$, $b=10$. The thickness $d_2$ is
variable: (a, b) $d_2=0.125$, (c, d) $d_2=0.127$, (e, f) $d_2=0.13$
$\mu$m.}
\end{figure}

\begin{figure}
\includegraphics[scale=0.75, clip=]{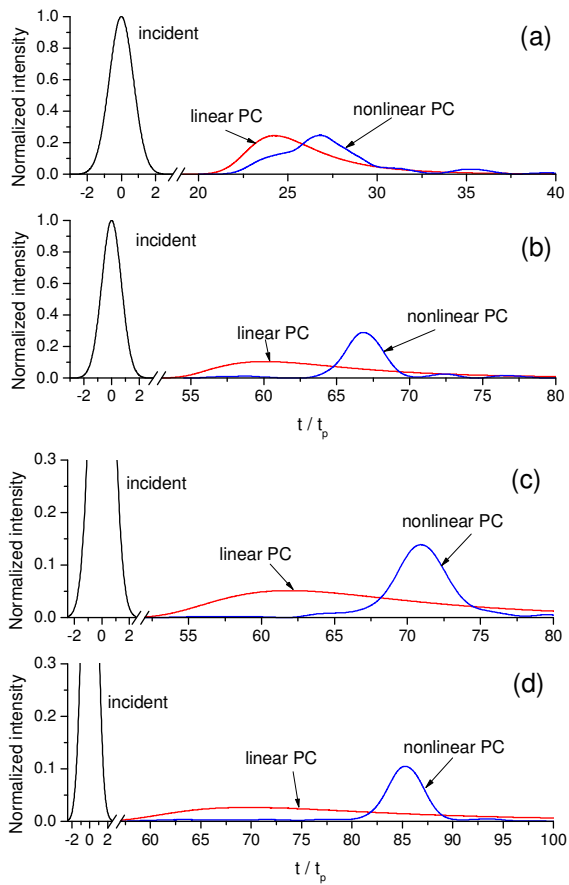}
\caption{\label{fig10} (Color online) The forms of transmitted
pulses after interaction with the photonic crystal with linear and
nonlinear layers $d_1$ ($n_1=1$, $b=10$). The number of periods is
(a) 100, (b, c, d) 250. The thickness $d_2$ is variable: (a, b)
$d_2=0.13$, (c) $d_2=0.127$, (d) $d_2=0.1285$ $\mu$m.}
\end{figure}

\begin{figure}
\includegraphics[scale=0.75, clip=]{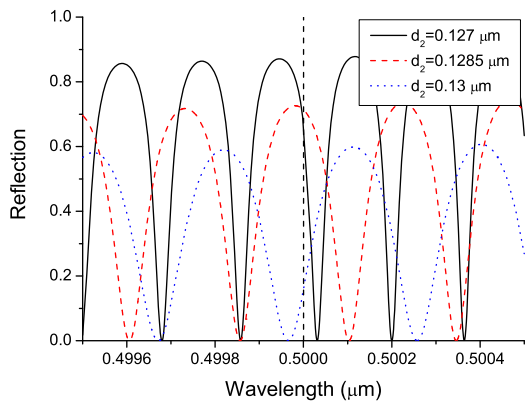}
\caption{\label{fig11} (Color online) Reflection spectra of photonic
crystal (number of periods 250) for different values of the
thickness $d_2$.}
\end{figure}

\begin{figure}
\includegraphics[scale=0.75, clip=]{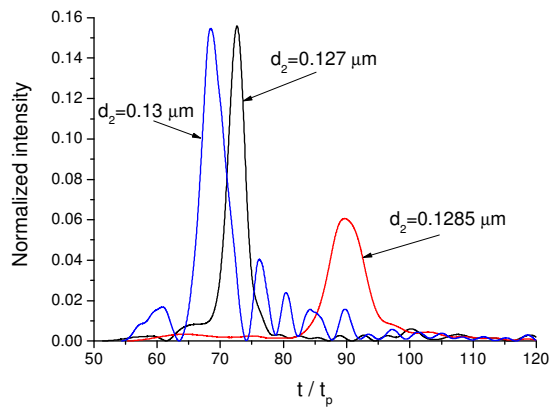}
\caption{\label{fig12} (Color online) The forms of transmitted
pulses after interaction with the photonic crystal (number of
periods 250) with both nonlinear layers ($n_1=1$, $n_2=3.5$, $b=10$)
for different thicknesses $d_2$.}
\end{figure}

As it was already stated, the changes of laser pulse characteristics
under propagation in nonlinear photonic band gap (PBG) structures
are of special interest in the present paper. The one-dimensional
photonic crystal to be considered here is a sequence of two
alternating layers with different thicknesses and background
refractive indices (further they are assumed to equal $n_1=1$ and
$n_2=3.5$). The thickness of the first layer is $d_1=0.4$ $\mu$m,
while the second one, $d_2$, is treated as a variable parameter
which allows to change spectral characteristics of the periodic
structure. The reflection spectra for different values of $d_2$ are
shown in Fig. \ref{fig8}a. They are plotted in the vicinity of the
main wavelength $\lambda=0.5$ $\mu$m. The pulse amplitude in this
section is considered to be equal to $\Omega_0=1.5\Delta\Omega_T$.

Light pulse propagation in photonic crystals, possessing strong
dispersion due to periodic variation of optical properties, results
in their spreading on large distances. As it is known \cite{Zhelt,
Vlasov}, for materials with sufficiently strong nonlinearity, the
competition between dispersion and nonlinear interaction leads to
effective compression of pulses. These results regard media with
Kerr-type nonlinearities. Here we consider some properties of the
systems combining PBG structure with the dense resonant medium (the
value of the NDD interaction constant is $b=10$).

The first such system is a structure containing a defect layer in
the middle of a sequence of alternating layers. The reflection
spectrum of such system with the linear defect (the refractive index
$n_3$ is unity) is shown in Fig. \ref{fig8}b. The well-known feature
of the photonic crystals with defects is seen: appearance of the
defect modes, i.e. narrow spectral peaks in the band-gap. Fig.
\ref{fig9} demonstrates the computation results of pulse
transmission and reflection for linear and nonlinear defect layers.
As transmission of the PBG structure increases (as $d_2$ changes),
the effect of nonlinear interaction on the pulse shape becomes
stronger leading to more effective compression (Fig. \ref{fig9}e).
As to reflected radiation, there are several pulses, the first one
being reflected directly from the photonic crystal layers. The
others were reflected after interaction with the dense resonant
medium.

It seems to be more interesting to consider a variant when the
layers of photonic crystal are nonlinear. Fig. \ref{fig10} gives the
simulation results when the layers $d_1$ of the PBG structure are
filled with the dense resonant medium. It is seen that use of
nonlinear layers allows to effectively compensate dispersive
spreading of pulse occurring in the case of linear photonic crystal.
Different values of the layer thickness $d_2$ correspond to
different values of reflectivity of the linear structure for the
radiation wavelength $\lambda=0.5$ $\mu$m (Fig. \ref{fig11}).
Naturally, more intensive transmitted pulse is observed for smaller
reflectivity (when $d_2=0.13$). At the same time, position of
wavelength with respect to PBG spectrum influences on the duration
of pulse transmission through the system: the smaller reflectivity
of the structure considered, the shorter time interval required for
the pulse to appear on its output. Comparing transmission durations
in Fig. \ref{fig10}a-c, one can note that radiation needs only about
$67t_p$ for $d_2=0.13$ $\mu$m, while for $d_2=0.1285$ $\mu$m (the
largest reflectivity) this pulse delay is equal to $85t_p$.

Finally, we consider another system when both layers of photonic
crystal are nonlinear. On the whole, the behavior of the system
(Fig. \ref{fig12}) remains approximately the same as in previous
case. The maximal change occurs for the pulse at $d_2=0.1285$
$\mu$m: the peak intensity decreases, while the time delay
increases. Note that at $d_2=0.13$ $\mu$m the effective splitting of
the pulse is observed. It may be connected with the change of
reflection characteristics of the nonlinear PBG structure under
consideration.

Thus, interaction of light pulse with nonlinear PBG structure leads
to effective compensation of dispersive spreading. On the other
hand, using of photonic crystals allows to control intensity and
time retardation of transmitted pulse. Additional possibilities to
control pulse properties are connected with use of a second one.

\section{\label{2pulse}Controlling pulse intensity by using another pulse}

\subsection{\label{coprop}Co-propagating pulses}

\begin{figure}
\includegraphics[scale=0.75, clip=]{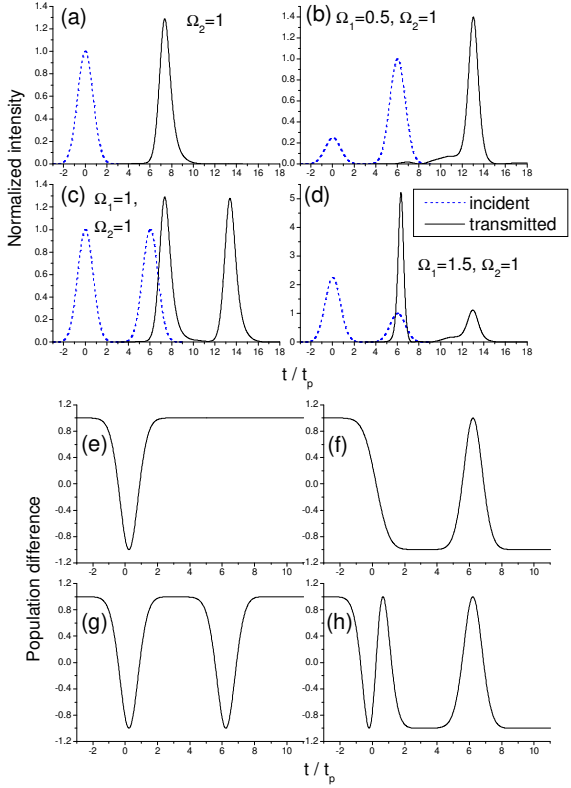}
\caption{\label{fig13} (Color online) Pulse forms (a-d) and
corresponding population difference dynamics on the entrance of the
medium (e-h). Intensity of the first pulse (b, f) $\Omega_1=0.5$,
(c, g) $\Omega_1=1$, (d, h) $\Omega_1=1.5$, while for the second one
$\Omega_2=1$. Pictures (a, e) are for the case of single pulse
$\Omega_2=1$. Other parameters: $L=100\lambda$, $b=10$.}
\end{figure}

\begin{figure}
\includegraphics[scale=0.75, clip=]{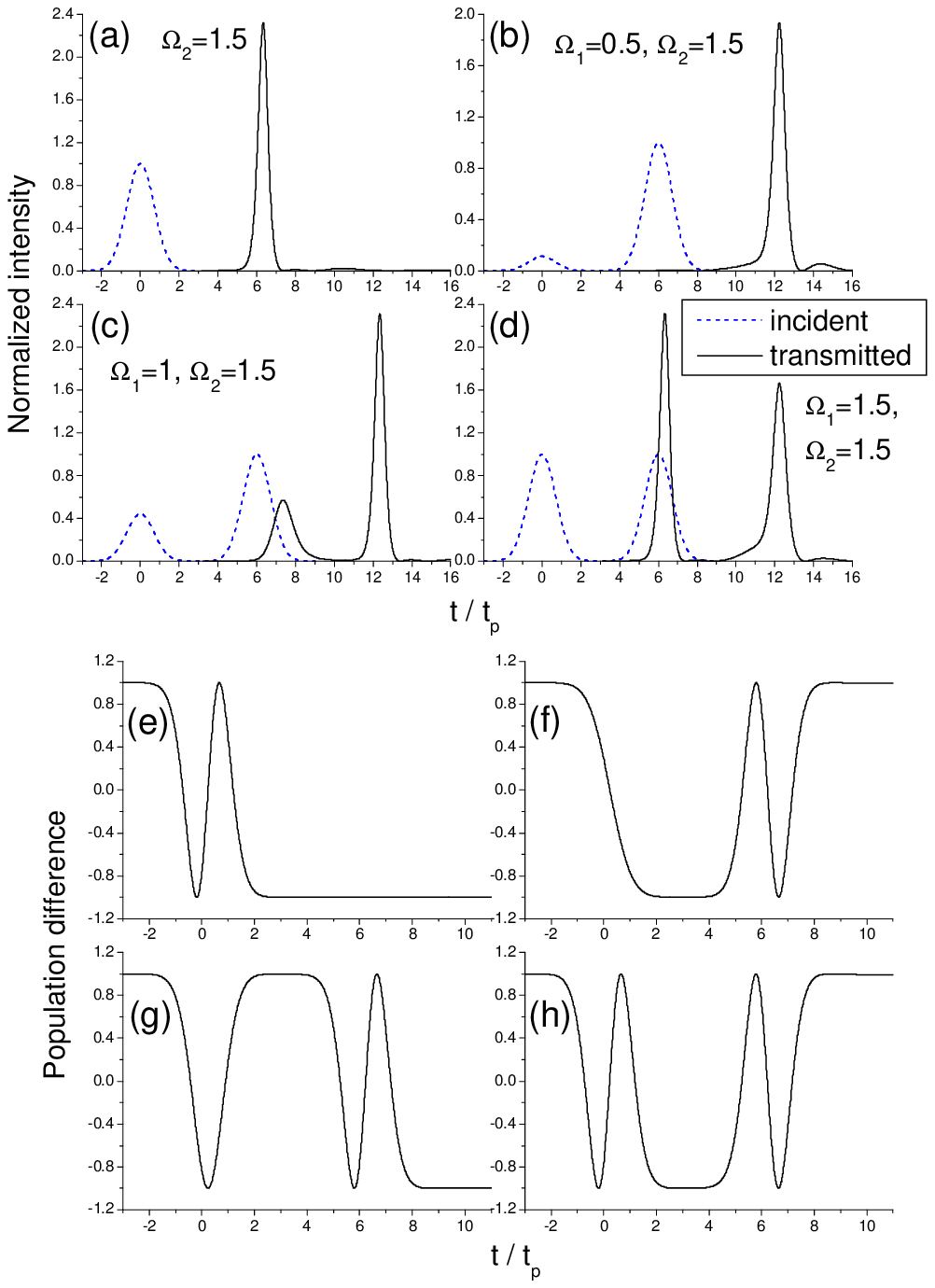}
\caption{\label{fig14} (Color online) Pulse forms (a-d) and
corresponding population difference dynamics on the entrance of the
medium (e-h). Intensity of the first pulse (b, f) $\Omega_1=0.5$,
(c, g) $\Omega_1=1$, (d, h) $\Omega_1=1.5$, while for the second one
$\Omega_2=1.5$. Pictures (a, e) are for the case of single pulse
$\Omega_2=1.5$. Other parameters: $L=100\lambda$, $b=10$.}
\end{figure}

\begin{figure}
\includegraphics[scale=0.75, clip=]{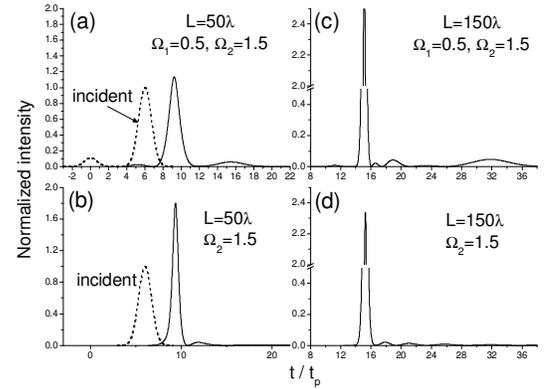}
\caption{\label{fig15} Pulse forms at different distances in medium.
Intensities of the pulses $\Omega_1=0.5$, $\Omega_2=1.5$. Pictures
(b, d) are for the case of single pulse $\Omega_2=1.5$. Parameter
$b=10$.}
\end{figure}

\begin{figure}
\includegraphics[scale=0.75, clip=]{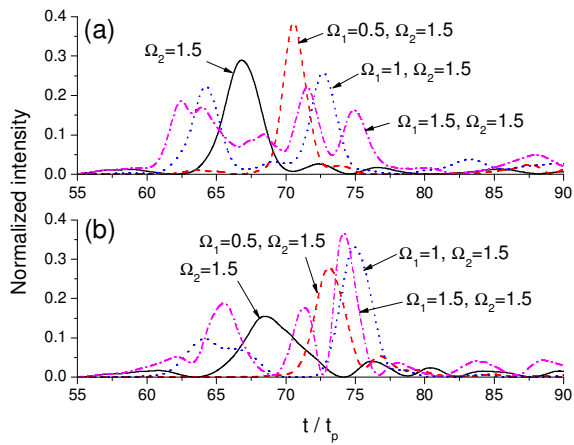}
\caption{\label{fig16} (Color online) The forms of transmitted
pulses after interaction with the photonic crystal (number of
periods 250) with (a) nonlinear layers $d_1$, (b) both nonlinear
layers. Intensity of the second pulse is $\Omega_2=1.5$; $\Omega_1$
is variable. Parameters: $n_1=1$, $n_2=3.5$, $d_1=0.4$ $\mu$m,
$d_2=0.13$ $\mu$m, $b=10$.}
\end{figure}

\begin{figure}
\includegraphics[scale=0.75, clip=]{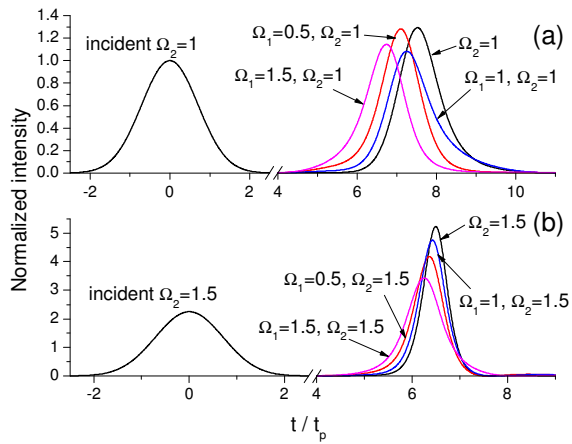}
\caption{\label{fig17} (Color online) The forms of transmitted pulse
$\Omega_2$ controlled with a counter-propagating pulse $\Omega_1$.
Intensity of the second pulse is (a, c) $\Omega_2=1$, (b, d)
$\Omega_2=1.5$; $\Omega_1$ is variable. Other parameters:
$L=100\lambda$, $b=10$.}
\end{figure}

\begin{figure}
\includegraphics[scale=0.75, clip=]{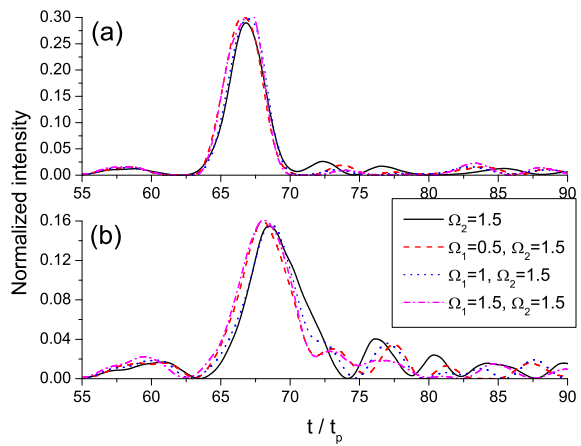}
\caption{\label{fig18} (Color online) The forms of transmitted pulse
$\Omega_2$ after interaction with the photonic crystal (number of
periods 250) with (a) nonlinear layers $d_1$, (b) both nonlinear
layers. Intensity $\Omega_2=1.5$; intensity of the
counter-propagating pulse $\Omega_1$ is variable. Parameters:
$n_1=1$, $n_2=3.5$, $d_1=0.4$ $\mu$m, $d_2=0.13$ $\mu$m, $b=10$.}
\end{figure}

Let us consider "one-by-one" propagation of two short pulses and
interaction between them via the dense resonant medium. As we have
seen in Section \ref{coher}, the population difference depends on
the amplitude of Gaussian pulse in periodic manner at realistic
values of parameter $b$. In this Section amplitudes will be
expressed in the units of this period $\Delta\Omega_T$. Hence, the
behavior of the second pulse with amplitude $\Omega_2$ differs
according to amplitude of the first one $\Omega_1$ and,
consequently, the state of the medium after it has passed. The
results of calculations are shown in Figs. \ref{fig13} and
\ref{fig14}. Obviously, if the first pulse returns the medium to the
ground state ($\Omega_1=1$), it practically does not effect on the
second one. But if after the passage of the first pulse the medium
is excited, the second pulse can demonstrate significant increasing
(Fig. \ref{fig13}b) or decreasing (Fig. \ref{fig14}b, d) of peak
intensity. One can treat this effect as controlling of the second
pulse intensity by using the first pulse.

The thickness of the dense resonant medium turned out to be an
important parameter (see Fig. \ref{fig15}). When the distance
traveled by the pulse $\Omega_2=1.5$ is $L=50\lambda$ or
$L=100\lambda$, the peak intensity of it appears to be greater in
the case of the single pulse than in two-pulse scheme. But at
$L=150\lambda$ (when the single pulse demonstrates attenuation due
to diffraction and dispersion) situation becomes reverse. In
general, it seems that the first pulse changes the distance of
optimal compression of the second one.

In order to make the efficiency of control higher, the PBG structure
can be used. For nonlinear layers $d_1$, more effective compression
is obtained only when $\Omega_1=0.5$ (Fig. \ref{fig16}a). However,
when both layers $d_1$ and $d_2$ are nonlinear, Fig. \ref{fig16}b
demonstrates increasing of peak intensities for all values of
$\Omega_1$. For example, for $\Omega_1=1.5$ it reaches approximately
twofold growth in comparison with the single pulse case. The reason
of this effect is that photonic crystal provides intense energy
exchange between the first and second pulses due to reflections on
the layers boundaries. The side effect is the difficulty of pulse
separation on the output of the system.

\subsection{\label{counterprop}Counter-propagating pulses}

Another scheme of controlling pulse intensity is connected with
utilizing of counter-propagating pulse. The advantage of this scheme
is the convenience of separation of pulses as they propagate in
opposite directions. The calculation results for the pulse
$\Omega_2$, which is controlled with the counter-propagating one
$\Omega_1$, are shown in Fig. \ref{fig17}. It is seen that change of
the intensity of the incident pulse $\Omega_1$ leads to the change
of the intensity of the transmitted pulse $\Omega_2$, in particular
to its decreasing in comparison with single pulse case. Since the
change of the length of optimal compression occurs in this scheme as
well, one can obtain increasing of the intensity of $\Omega_2=1.5$
pulse on the distance of about $200\lambda$ when $\Omega_1=1$.

Fig. \ref{fig18} shows that the interaction between pulses in the
nonlinear photonic crystal is negligible, in contrast to the case of
co-propagating pulses. It seems to be the result of weak coupling of
the pulses in such system, as they propagate almost independently.
This is the great disadvantage in the view of controlling
possibilities, but it can be used for simultaneous work with two
pulses moving in opposite directions.

\section{\label{conc}Conclusion}

The consideration of intensive radiation interaction with the dense
resonant medium in the case of coherent pulse regime allows to
conclude that the influence of near dipole-dipole interactions can
be neglected at realistic parameters of the medium, in contrast to
stationary case. At the same time, one can observe pulse compression
(and splitting) which has certain optimal distance due to the
processes of diffraction and dispersion. This property of nonlinear
compression can be used to compensate dispersive spreading in
photonic band gap structure containing the resonant medium
considered.

It turned out that dense resonant medium allows to control pulse
with another pulse in schemes of co- and counter-propagating pulses.
It seems to be useful from the point of view of prospective
techniques of optical information storage and processing. Photonic
crystal makes the process of controlling more effective, at least in
the case of co-propagating pulses. For counter-propagating ones, it
appears not to be efficient, nevertheless independence of pulse
propagation in opposite directions can be used for parallel work
with two pulses.

\end{document}